\documentclass[prl,aps,twocolumn,groupedaddress,showpacs,floatfix]{revtex4}
\usepackage{amsmath,amssymb,multirow,epsfig,bm}

\newcommand{\etal}{\textit{et al.,\;}}
\newcommand{\beq}{\begin{equation}}
\newcommand{\eeq}{\end{equation}}
\newcommand{\bea}{\begin{eqnarray}}
\newcommand{\eea}{\end{eqnarray}}

\newcommand{\benn}{\begin{displaymath}}
\newcommand{\eenn}{\end{displaymath}}
\newcommand{\intrange}{r_{0}}                          
\newcommand{\sclgth}{a}                                
\newcommand{\reff}{r_{\text{ef\mbox{}f}}}              
\newcommand{\dens}{\rho}                               
\newcommand{\kF}{k_{F}}                                
\newcommand{\eF}{\varepsilon_{F}}                      
\newcommand{\fm}{\,\text{fm}}                          
\newcommand{\MeV}{\,\text{MeV}}                        
\renewcommand{\vec}[1]{\textbf{#1}}                    
\newcommand{\vecp}[1]{\textbf{#1}^{\prime}}            
\newcommand{\pcut}{p_{\text{cut}}}                     
\newcommand{\Ns}{N_{s}}                                
\newcommand{\EFFG}{E_{\text{FFG}}}                     
\newcommand{\Tc}{T_{c}}                                

\begin{document}
\title{\mbox{Quantum Monte Carlo} study of dilute neutron matter at finite temperatures}

\author{Gabriel Wlaz\l{}owski and Piotr Magierski }
\affiliation{Faculty of Physics, Warsaw University of Technology,
ulica Koszykowa 75, 00-662 Warsaw, POLAND }

\begin{abstract}
We report results of fully non-perturbative, Path Integral Monte Carlo (PIMC) calculations for dilute neutron matter.
The neutron-neutron interaction in the $s$ channel is parameterized by the scattering length and the effective range.
We calculate the energy and the chemical potential as a function of temperature
at the density $\dens=0.003\fm^{-3}$. The critical temperature $\Tc$ for
the superfluid-normal phase transition is estimated from the finite size scaling of the condensate fraction.
At low temperatures we extract the spectral weight function $A(p,\omega)$ from the imaginary time propagator using the methods of maximum entropy and singular value decomposition. We determine the quasiparticle spectrum, which can be accurately parameterized by three parameters: an effective mass $m^*$, a mean-field potential $U$, and a gap $\Delta$. 
Large value of $\Delta/\Tc$ indicates that the system is not a BCS-type superfluid at low temperatures.
\end{abstract}

\date{\today}

\maketitle
Dilute neutron matter is one of the simplest many-body nuclear systems.
At sufficiently small densities its properties originate from the two-body $s$-wave interaction only.
It is known that neutron matter has a positive pressure at all densities (contrary to nuclear matter) which
prevents fragmentation and it becomes superfluid at low temperatures.
From the theoretical point of view, pure and dilute neutron matter is a fascinating system 
since at a certain density range it becomes a nearly-universal Fermi gas.
Such systems are presently of great interest as a result of an extraordinary progress in the field of cold atoms 
which have taken place over the last few years and in fact opened new chapter in many-body physics 
(see \cite{reviews} and references therein). 
Taking advantage of the Feshbach resonances experimentalists can control the strength of the 
atom-atom interaction and achieve the so-called unitary regime. It corresponds to the situation where the average 
distance between fermionic atoms is large as compared to the interaction range $\intrange$, but much smaller than
the scattering length $\sclgth$ ie. \mbox{$\dens\intrange^{3}\ll 1 \ll \dens|\sclgth|^{3}$}, 
where $\dens$ is the particle number density. In the unitary regime the properties of dilute Fermi gases are universal, 
independent of the details of the interaction. 
Universality of these systems make them fascinating theoretical playground, and obtained results 
turned out to be relevant to a wide range of fields like string theories, the quark-gluon plasma, and high $T_c$ superconductors. 

Since the $^{1}S_{0}$ neutron-neutron interaction
is characterized by the large scattering length $\sclgth\approx-18.5\fm$, 
the unitary regime can be thought of as a limiting case of dilute neutron matter
at the density range varying from $0.001$ to $0.01\fm^{-3}$.
One has to remember, however, that the influence of the effective range ($\reff\approx 2.8\fm$) cannot be 
ignored since $k_{F}\reff$ is of the order of unity \cite{SchwenkPethick}.
The importance of other channels as well as of three-body forces is increasing
with density. However at the density $0.003\fm^{-3}$, which we study in this paper,
their influence is marginal as compared to uncertainties of PIMC method
and therefore will be neglected \cite{BaldoMaieron,GezerlisCarlson_new}. 

Since even for the density $\dens=0.001\fm^{-3}$
dilute neutron matter is a strongly correlated Fermi gas ($|\kF\sclgth| \gg 1$)
only non-perturbative approaches are able to gain reliable insight into physics of this system.
The large class of such methods, which are known under the general name of Quantum Monte Carlo (QMC), have
been used to date, although most of them concern the zero temperature properties 
\cite{FantoniEtAl,GandolfiEtAl,EpelbaumEtAl}. 
The finite temperature behavior has been studied in \cite{MullerEtAl}.
This work presents the first {\it ab initio}, fully non-perturbative evaluation of thermal properties of 
low-density neutron matter (at about 2\% of nuclear saturation density) 
free of uncontrolled approximations within PIMC method. 
We focus on the effects generated by the finite effective range.

Contrary to cold atomic gases, in order to capture physics of dilute neutron matter 
one has to use more realistic interaction than a simple contact, 
delta-like force. In the present paper we employ the two-body potential of the form:
\begin{equation}\label{eqn:interaction}
 V(\vec{r}-\vecp{r})=\left\lbrace \begin{array}{ll}
                      6g,&\vec{r}-\vecp{r}=0\\
                      g,&\vec{r}-\vecp{r}\in \mathcal{N}_{b}\\
                      0, &otherwise
                     \end{array}\right.  , 
\end{equation}
where $\mathcal{N}_{b}=\{(\pm b,0,0),(0,\pm b,0),(0,0,\pm b) \}$ represents the set of the 
nearest neighbor coordinates. 
This particular form of the interaction is especially designed for the cubic lattice
with the lattice constant $b$ and enables to construct 
a fully non-perturbative approach without the
sign problem (for more details see Ref. \cite{WlazlowskiMagierski}). 
It depends on two parameters ($g$ and $b$) which are adjusted to correctly reproduce 
the scattering length and the effective range of neutron-neutron $^{1}S_{0}$ scattering amplitude \cite{WlazlowskiMagierski2}. 
Hence we consider the system on a 3D spatial cubic lattice of length $L=\Ns b$
with periodic boundary conditions. The lattice spacing $b$ and size $L$ introduce 
the natural ultraviolet (UV) and infrared (IR) momentum cut-offs given by 
$\pcut=\pi/b$ and $p_0=2\pi/L$, respectively. 
The momentum space has the shape of a cubic lattice, with size $2\pi/b$ and spacing $2\pi/L$.
To simplify the analysis, however, we place the spherically symmetric UV cut-off, 
including momenta $p \le \pcut $.  

To evaluate numerically expectation values of observables we have followed
the path integral approach described in Ref. \cite{bcs-bec}.
Using Trotter expansion and
subsequently Hubbard-Stratonovich (H-S) transformation,
the evaluation of the emerging path integral 
was performed using the Metropolis importance sampling. 
The crucial modification of the procedure described in \cite{bcs-bec} consists in
the construction of such H-S transformation which allow to incorporate 
the off-site part of the interaction without generation of the sign problem.
Namely, we have used the discrete H-S transformation of the form \cite{WlazlowskiMagierski}: 
\begin{equation}\label{eqn:hs-transformation} 
 e^{-\tau\hat{V}}=\prod_{\vec{r}-\vecp{r}\in\mathcal{N}_{b}}\prod_{\lambda=\uparrow\downarrow}\dfrac{1}{k}\sum_{i=1}^{k}e^{\sigma_{i}(\vec{r},\vecp{r})[\hat{n}_{\lambda}(\vec{r})+\hat{n}_{\lambda}(\vecp{r})]},
\end{equation}
where $\sigma_{i}$ are real numbers and $\hat{n}_{\lambda}(\vec{r})$ is the occupation number operator. 
The notable feature of this H-S transformation is the time reversal invariance
of the corresponding imaginary time evolution operator. 
This property ensures that the probability measure used in the Metropolis algorithm is always positive 
\cite{WlazlowskiMagierski,KooninEtAl}.

Calculations were performed on the lattice of size $\Ns=8$ with the lattice constant
$b=3.21\fm$. The chemical potential was chosen in such a way to keep the total number of particles between 53 and 57,
which corresponds to the density $\kF\simeq 0.45\fm^{-1}$. 
The temperatures span the interval from $0.06\,\eF$ ($0.26\MeV$) to $1.0\,\eF$
 ($4.3\MeV$), where $\eF$ is the Fermi energy. The number of imaginary time steps required to reach
the convergence of the algorithm varies with temperature. 
At the lowest temperature $2360$ imaginary time steps have been applied, 
whereas for the highest temperature only $216$. 
The kinetic energy part of the Hamiltonian is defined in the restricted momentum space 
($p \le \pcut$) using the dispersion relation of the form $\varepsilon(\vec{p}) = p^{2}/2m$.
Consequently during the imaginary time evolution 
the FFT algorithm has been used to switch between momentum and coordinate spaces \cite{bcs-bec}.
The number of generated uncorrelated Monte Carlo samples
allows to decrease the statistical error below 5\%.  
At low temperatures the Singular Value Decomposition technique was applied to avoid instabilities 
of the algorithm. In all runs the single-particle occupation probabilities for the highest energy states were 
below one percent at all temperatures. We have also performed a few exploratory simulations for the lattice of size 
$\Ns=10$. The results were in a good agreement with those for $\Ns=8$ lattice.

\begin{figure}[tb]
\includegraphics[width=8.00cm]{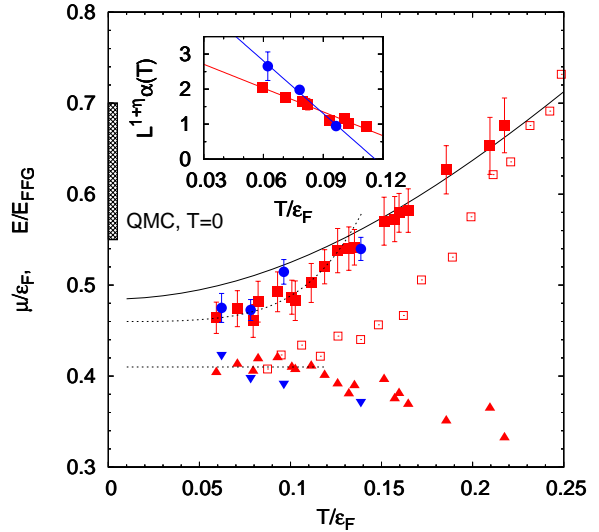}
\caption{ (Color online) Total energy $E$ and chemical potential $\mu$ 
as a function of temperature for dilute neutron matter at the density $\rho=0.003\fm^{-3}$ ($\kF\simeq 0.45\fm^{-1}$). 
The total energy is denoted by 
red squares ($8^{3}$ lattice) and blue circles ($10^{3}$ lattice). 
The red up triangles and blue down triangles correspond to the chemical potential for $8^{3}$ and $10^{3}$ lattice, respectively.
The solid line represents the energy of the noninteracting Fermi gas, shifted by a constant value. 
The dashed line shows an extrapolation of the energy and the chemical potential to $T=0$ limit. For comparison the total energy of the unitary Fermi gas is also plotted (open red squares). Dashed area for $T=0$ denotes the range where the results of other 
QMC results are located (see for example Ref. \cite{GandolfiEtAl}).
In the inset the rescaled condensate fraction is shown as a function of temperature, 
red squares and blue circles denote $8^3$ and $10^{3}$ lattices, respectively. 
Crosspoint determines the critical temperature of the superfluid-normal phase transition, $\Tc\approx0.09\,\eF$.
\label{fig:eos} }
\end{figure}
In the Fig. \ref{fig:eos} the low temperature behavior of the total energy and the chemical potential 
is presented for two different lattice sizes. 
The (shifted) total energy versus temperature for the free Fermi gas at the same particle density has also been plotted 
(solid line). Note that after shifting of the free Fermi gas energy by $0.52\,\EFFG$ the curve
reproduces Monte Carlo results
for $T>0.15\,\eF$ ($\EFFG=\frac{3}{5}N\eF$ is the free Fermi gas energy at $T=0$). 
Below this temperature the deviation from the free Fermi gas behavior is clearly visible.
The chemical potential is approximately constant for  $T<0.1\,\eF$. 

The critical temperature of the superfluid-normal phase transition has been determined
using the method based on the finite size scaling of the correlation function. 
Similar technique was used to determine the critical temperature at the unitary limit 
(see Refs. \cite{bcs-bec,BurovskiEtAl} for details). 
The volume-dependent estimation of the critical temperature $\Tc^{(ij)}$ was obtained by finding the crossing point of the 
rescaled condensate fraction for two different lattice sizes $N_{i,j}$. As $N_{i,j}\rightarrow\infty$, the series 
$\Tc^{(ij)}$ converges to $\Tc$ and one can extract the limiting value. 
We have determined
$\Tc$ using results for two lattices $N_{i,j}=8,10$. Such large lattices and rather small filling factor which in both cases reads $\nu=N/2\Ns^{3}\approx5\%$ are enough to estimate the critical temperature with uncertainty smaller than $20\%$
(in fact this procedure applied to the unitary gas gives estimation of the $\Tc$ with the relative error smaller than $10\%$).
The estimate of the critical temperature reads $\Tc\approx 
0.09\,\eF$. Note that $\Tc$ is considerably lower than the temperature for the onset of deviation 
from the free Fermi gas behavior.

Within the PIMC framework one cannot reach directly the $T=0$ limit. 
However the ground state energy can be obtained by performing an extrapolation of results to zero temperature limit. 
In our case this procedure provides the ground state energy $E/\EFFG=0.46(2)$ ($E/N=1.22(5)\MeV$).
This value is considerably lower (by about 20\%-40\%) than values
obtained by other MC calculations
(see for example Ref. \cite{GandolfiEtAl}). This is most likely due to the fact, that our approach 
is based on fully unrestricted path integral calculations and, within statistical errors 
due to the Monte Carlo procedure, gives essentially exact results.

The gap in the fermionic spectrum, related to superfluidity,  
has been computed from the spectral weight function $A(\vec{p},\omega)$ by performing  
the analytic continuation of the imaginary time propagator ${\cal G}(\vec{p},\tau)$ 
to real frequencies \cite{fw}.
This procedure is equivalent to solving the integral equation:
\begin{equation}
{\cal G }(\vec{p},\tau)=-\frac{1}{2\pi}\int_{-\infty}^{+\infty}
d\omega A(\vec{p},\omega)\frac{\exp(-\omega\tau)}{1+\exp(-\omega\beta)},
\label{eqn:Ap}
\end{equation} 
where ${\cal G }(\vec{p},\tau)$ is known from the Monte Carlo calculations for $51$ different values of $\tau\in[0,\beta=1/T]$.
The inverse problem is however numerically ill-posed i.e. there is an infinite class of solutions for $A(\vec{p},\omega)$ which satisfy Eq. (\ref{eqn:Ap}) within uncertainties generated by the Monte Carlo method. Therefore we have used two independent methods based on completely different mathematical approaches. 

The first one, the maximum entropy method, is based on Bayes' theorem \cite{jaynes}. It 
treats the values of ${\cal \tilde{G}}(\vec{p},\tau_{i})$ ($i=0,1,...,50$) provided by QMC simulation as normally 
distributed random numbers, around the true values ${\cal G}(\vec{p},\tau_{i})$, and searches for 
the most probable solution assuming some \textit{a~priori} knowledge concerning the spectral function. 
As an \textit{a~priori} information we have used constraints:
\begin{eqnarray}
& & A(\vec{p},\omega) \ge 0, \quad \quad
\int_{-\infty}^{+\infty}\frac{d\omega}{2\pi} A(\vec{p},\omega) = 1,  \label{eqn:Ap_con1} \\
& & \int_{-\infty}^{+\infty}\frac{d\omega}{2\pi} A(\vec{p},\omega)\frac{1}{1+\exp(\omega\beta)} = n(\vec{p}),
\label{eqn:Ap_con2}
\end{eqnarray}
and we have assumed a Gaussian-like structure for $A(\vec{p},\omega)$. 
In the formula (\ref{eqn:Ap_con2}) $n(\vec{p})$ represents the occupation probability of the state 
with momentum~$\vec{p}$ which is known from the Monte Carlo simulation.

The second method is based on the singular value decomposition (SVD) of
the integral kernel ${\cal K}$ of Eq.~(\ref{eqn:Ap}), which can be rewritten in the
operator form as ${\cal G}(\vec{p},\tau_{i})=({\cal K}A)(\vec{p},\tau_{i})$. The operator ${\cal K}$ possesses the singular system which forms a suitable basis for the expansion of the projected spectral weight function $\tilde{A}(\vec{p},\omega)$ onto a  subspace where the inverse problem is well-posed \cite{svd1}. 
Since the method provides only projection of the ``true'' 
solution, it does not require any \textit{a~priori} information, contrary to the maximum entropy method. 
However, since 
${\cal G}(\vec{p},\tau_{i})$ include statistical errors
due to the Monte Carlo procedure, the projected solution $\tilde{A}(\vec{p},\omega)$ is also 
affected by this uncertainty. 
One can use this flexibility by choosing the solution satisfying the constraints~(\ref{eqn:Ap_con1})~\cite{svd2}.
The details of both methods will be discussed elsewhere \cite{mw}. 

\begin{figure}[tb]
\includegraphics[width=9.0cm]{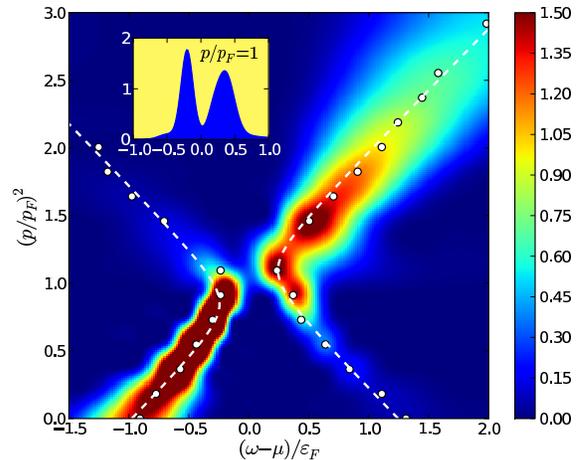}
\caption{ (Color online) Spectral weight function $A(p,\omega)$ at the temperature $T=0.06\,\eF$ and lattice size $\Ns=10$ obtained by the maximum entropy method. Points indicate localizations of maxima for fixed values of momenta. Dashed lines correspond to the fit of the BCS-type formula given by (\ref{eqn:Ep}). In the inset the spectral weight function at the Fermi level is presented.
\label{fig:A76} }
\end{figure}

The spectral weight function for the lowest temperature $T=0.06\,\eF$ obtained for $\Ns=10$ lattice is shown in 
the Fig. \ref{fig:A76}. The same outcome has been
generated by both methods (maximum entropy and SVD) independently. 
The presence of a ``pairing'' gap is clearly visible for this temperature. 

\begin{figure}[tb]
\includegraphics[width=5.8cm,angle=-90]{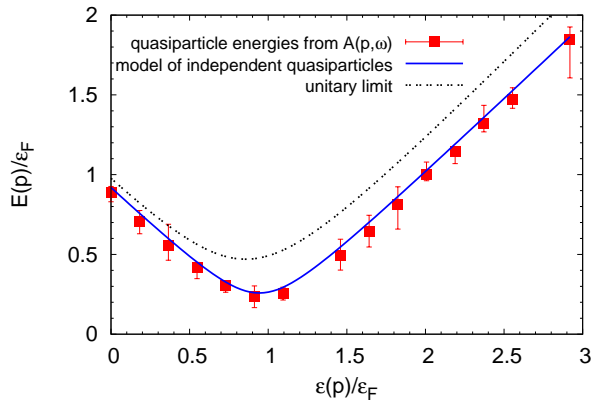}
\caption{ (Color online) Quasiparticle energies $E(p)$ (squares) extracted
from the spectral weight function $A(p,\omega)$ at $T=0.06\,\eF$.
The line denotes results obtained 
under assumption that the system is composed of independent quasiparticles. Dashed line corresponds to 
quasiparticle energies at the unitary limit.
\label{fig:E76} }
\end{figure}

Figure \ref{fig:E76} presents the quasiparticle excitation spectrum extracted from the spectral weight function for 
$T=0.06\,\eF$.
We have found that the quasiparticle excitations can be accurately parameterized by the BCS-like formula:
\begin{equation}
E(\vec{p}) = \pm \sqrt{ \left ( \frac{p^2}{2m^*}-\mu +U \right )^2+\Delta^2},
\label{eqn:Ep}
\end{equation} 
where $m^{*}$ is an effective mass, $U$ the mean field potential and $\Delta$ is the ``pairing'' gap. The values of these 
parameters were estimated as $m^{*}/m=1.1(1)$, $U/\eF=-0.26(6)$ and  $\Delta/\eF=0.25(5)$. 

Note that the ratio $\Delta/\Tc\approx 2.8$ is significantly higher than the well-known value $1.76$ predicted by BCS theory. 
The similar deviation from the BCS value is typical for high-temperature superconductors \cite{htc} 
and also for cold atomic gases in the unitary regime \cite{bcs-bec}.
Therefore we conclude, that the dilute neutron matter at this density is not a BCS-type superfluid. Note also that to estimate the value of $\Delta/\Tc$ we have used the value of the energy gap at the temperature $T=0.06\,\eF$, which is expected to be slightly lower than the value of the gap at zero temperature.

It is instructive to compare quasiparticle excitation energies with those extracted from the susceptibility function
under assumption that the system is composed of independent quasiparticles. Under this assumption the imaginary time propagator is simply given by:
\begin{equation}
{\cal G}(\vec{p},\tau)=-\dfrac{e^{-\tau E(\vec{p})}}{1+e^{-\beta E(\vec{p})}},
\label{eqn:ImaginaryTimePropagatorIQP}
\end{equation}  
and one can easily evaluate the susceptibility:
\begin{equation}
 \chi(\vec{p})=-\int_{0}^{\beta}d\tau\,{\cal G}(\vec{p},\tau)
	=\dfrac{1}{E(\vec{p})}\dfrac{e^{\beta E(\vec{p})}-1}{e^{\beta E(\vec{p})}+1}.
\label{eqn:ResponseFunctionIQP}
\end{equation}
From the calculated one-body propagator within the Monte Carlo algorithm one can
extract the spectrum of the elementary fermionic excitations inverting the Eq. (\ref{eqn:ResponseFunctionIQP}).
The extracted spectrum of quasiparticle energies turns out to reproduce very well (within error bars) the quasiparticle spectrum derived from the spectral function, see Fig.~\ref{fig:E76}. The same property is shared by unitary cold atomic gas 
at temperatures below the critical temperature \cite{MagierskiEtAl}.

Comparison of our results with those obtained in the limit $\reff\rightarrow 0$ provides an information 
about the influence of the effective range. From the data reported in Ref. \cite{bcs-bec} we infer that the effects of the 
effective range do not significantly alter the ground state energy. 
The value of the energy gap and the critical temperature
decreases considerably (at $\reff\rightarrow 0$: $\Delta^{(0)}/\eF\approx 0.41$ and 
$\Tc^{(0)}/\eF\approx 0.13$). 
However, surprisingly the ratio $\Delta^{(0)}/\Tc^{(0)}\approx 3.2$ remains
approximately constant (taking into account uncertainties of our estimation) when increasing $\reff$ to the value associated
with $^{1}S_{0}$ neutron-neutron interaction. 
Note also that the equation of state exhibits the existence of the second temperature scale, 
which can be attributed to the onset of deviations of $E/\EFFG$ from the (shifted) energy of 
the free Fermi gas. 
It bears similarity to the case of the unitary Fermi gas, where the existence of the 
so-called ``pseudogap'' above $\Tc$ is reported \cite{MagierskiEtAl}.

Summarizing, our results do not indicate 
the presence of qualitative changes in comparison to the case of zero effective range.
In conclusion the main aspects of physics 
at the unitary regime survive in the limit of dilute neutron matter.

We thank Aurel Bulgac for discussions.
Support from the Polish Ministry of Science under contracts No. N N202 328234, N N202 128439 
and by the UNEDF SciDAC Collaboration under DOE grant DE-FC02-07ER41457
is gratefully acknowledged. Use of computers at the Interdisciplinary Centre for Mathematical and Computational Modelling (ICM) at Warsaw University is also gratefully acknowledged.


\end{document}